%% file: main.tex
\documentclass[sigconf,nonacm]{acmart}

\makeatletter
\@ACM@balancefalse
\makeatother

\AtBeginDocument{%
  }

\setcopyright{none}
\usepackage{graphicx}
\usepackage{amsmath}
\usepackage{booktabs}
\usepackage{colortbl}
\definecolor{riskshade}{HTML}{B64342}
\DeclareRobustCommand{\rev}[1]{#1}
\usepackage{placeins}
\usepackage{balance}
\usepackage{algorithmic}
\usepackage{algorithm}


\input{math_commands}

\title{Right Family, Wrong Skill: Evaluating Risk Exposure in Agent Skill Retrieval}

\author{Jiandong Ding}
\affiliation{%
  \institution{Huawei Technologies Co., Ltd.}
  \city{Shanghai}
  \country{China}}
\email{dingjiandong2@huawei.com}

\author{Honglei Ji}
\affiliation{%
  \department{Department of Obstetrics}
  \institution{Shanghai East Hospital, Tongji University School of Medicine}
  \city{Shanghai}
  \postcode{200123}
  \country{China}}
\email{hongleijish@163.com}

\author{Ming Liu}
\affiliation{%
  \department{Department of Obstetrics}
  \institution{Shanghai East Hospital, Tongji University School of Medicine}
  \city{Shanghai}
  \postcode{200123}
  \country{China}}
\email{lmliuming@tongji.edu.cn}

\author{Tao Duan}
\affiliation{%
  \department{Department of Obstetrics}
  \institution{Shanghai East Hospital, Tongji University School of Medicine}
  \city{Shanghai}
  \postcode{200123}
  \country{China}}
\email{drduantao@126.com}
\renewcommand{\shortauthors}{Ding et al.}

\begin{document}

\begin{abstract}
\input{sections/0_abstract}
\end{abstract}


\maketitle

\begin{figure}[!t]
\centering
\includegraphics[width=\columnwidth]{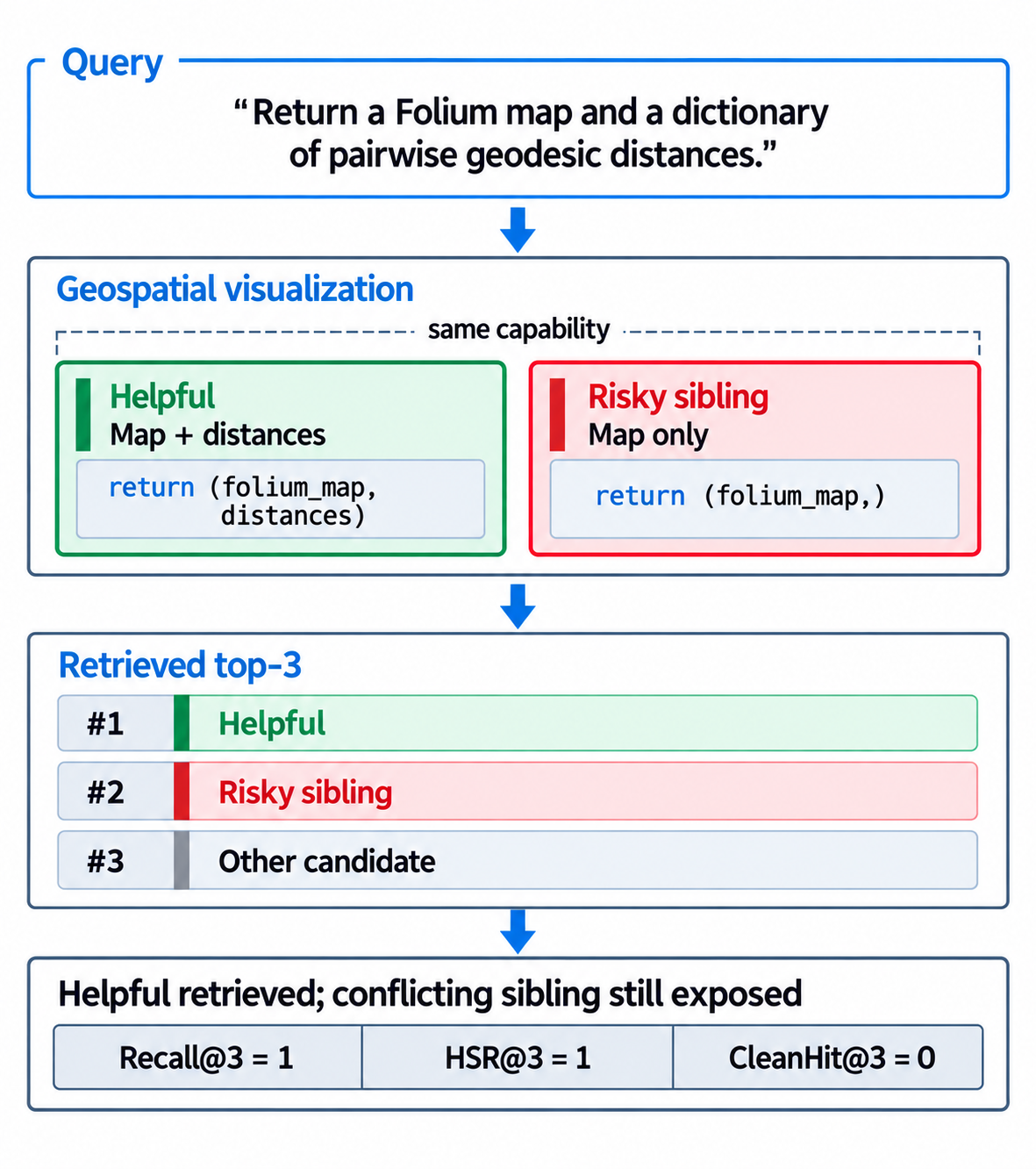}
\caption{A same-capability return-contract conflict. The query requires both a map
and distances; the sibling omits the distances from its return value.
\rev{Although the helpful skill ranks first, its conflicting sibling remains
in the list: Recall@3 = 1, HSR@3 = 1, and CleanHit@3 = 0.}}
\Description{A query asks for a Folium map and distance dictionary. Two
same-capability skills differ only in their return contract. An illustrative
top-three list contains both, yielding recall and HSR of one and CleanHit of zero.}
\label{fig:skillresolve-concept}
\end{figure}

\input{sections/1_introduction}

\begin{figure*}[t]
\centering
\includegraphics[width=\textwidth]{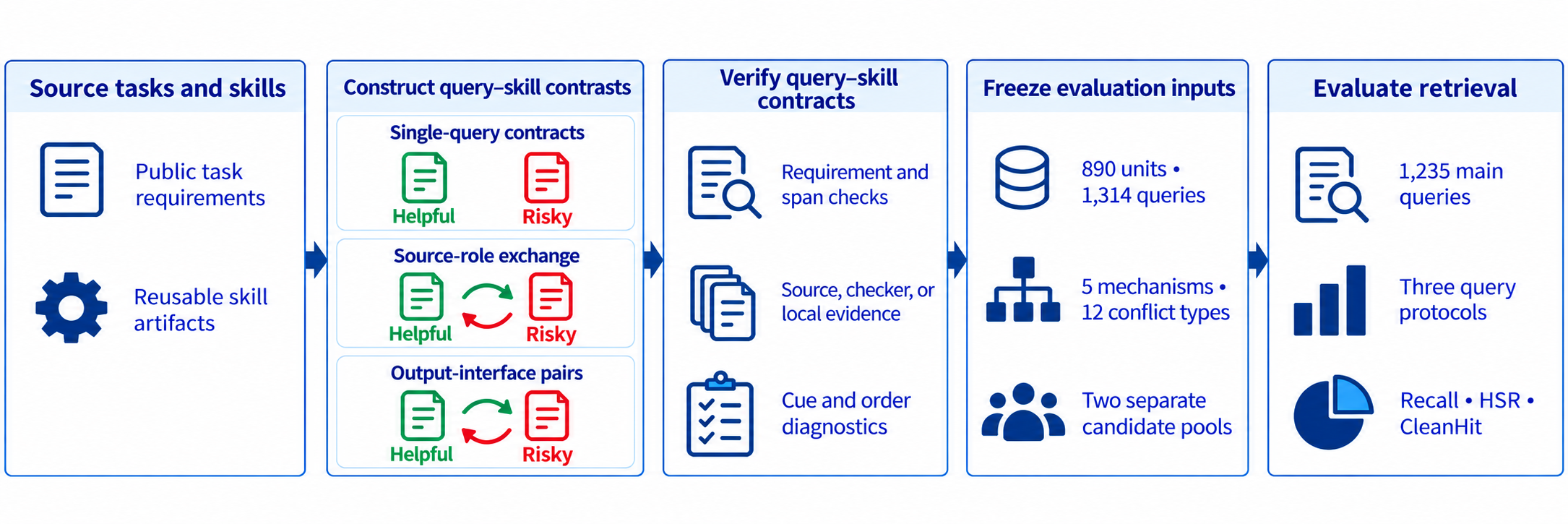}
\caption{Construction and evaluation of SameCapRisk-Bench. Single-query units
bind a source requirement to a conflicting sibling. Two paired protocols
reverse representative roles through source-evidence or exact output-interface
requests. Each contrast is checked against its recorded query and skill
texts before evaluation in two separate fixed pools.}
\Description{A left-to-right flow shows source tasks and skills, three
query protocols, checks against the recorded query and skill texts, separate
candidate pools, and retrieval evaluation.}
\label{fig:benchmark-pipeline}
\end{figure*}

\input{sections/2_related_work}
\input{sections/3_problem_method}
\input{sections/4_experiments_working890}
\input{sections/5_analysis_limitations_working890}
\input{sections/6_conclusion}

\bibliographystyle{ACM-Reference-Format}
\bibliography{references}

\appendix
\input{sections/A_appendix_working890}

\end{document}

%% file: math_commands.tex
\newcommand{\cC}{\mathcal{C}}
\newcommand{\cD}{\mathcal{D}}

\newcommand{\query}{q}

\newcommand{\spos}{s^{+}}
\newcommand{\sneg}{s^{-}}

%% file: sections/0_abstract.tex
A skill can match a task's topic while conflicting with its resource,
procedure, or output requirements. We study this as same-capability
risk-exposure retrieval and introduce \textbf{SameCapRisk-Bench}: 890 units
and 1,314 query cases across five mechanisms and twelve conflict types.
Each unit pairs a skill that meets a query requirement with a
same-capability skill that violates it, with evidence for that distinction.
The evaluation tests source-task requirements and two kinds of paired
queries that reverse which skill fits: changing the requested evidence
role or exact output interface while holding the skills fixed.
To capture both retrieval success and conflicting exposure, Recall tracks
helpful hits, harmful sibling rate (HSR) tracks exposure of the conflicting
sibling, and CleanHit requires a helpful hit without that exposure.

Four public skill retrievers expose conflicting siblings at HSR@3 of
0.737--0.881 on source-task contracts, compared with 0.099--0.12 on
controlled source-role queries. Across the fixed mixture of 1,235 held-out
queries, their Recall@3 is 0.903--0.944 and HSR@3 is 0.344--0.393.
The conflicting sibling ranks first on 24.8--29.2\% of source-task queries
for these four retrievers. We also examine what the reranker receives:
truncating long skills can remove the passages where the two skills'
contracts differ. With the same BGE top-20 candidates, increasing the
reranker's input budget from 512 to 4,096 tokens reduces source-task
sibling-first errors by 8.27~pp, with an uncertain
CleanHit gain.

%% file: sections/1_introduction.tex
\section{Introduction}

Agent skills package instructions, scripts, and resources for reuse across
tasks \citep{registryrepository2026,skillsnotislands2026}. For installed skill
catalogs, progressive disclosure exposes names and descriptions first,
then loads instructions and supporting resources as needed
\citep{agentskillsintegration2026}. Larger libraries motivate a retrieval
stage that narrows the candidates for selection \citep{skillret2026}.
SkillRouter, for example, combines dense retrieval with cross-encoder
reranking \citep{skillrouter2026}. Retrieval and on-demand loading can work
together: a system can first shortlist skills, then load the selected
instructions. We study the retrieval and ranking stage, which shapes the
skills available for subsequent selection and use.

Within a retrieved list, skills with similar capabilities may rely on
different resources, preconditions, or output interfaces. These differences
matter when a query requires a particular behavior. We ask: when two skills
serve the same broad purpose, which one meets the current query's requirements?

Suppose a query asks for a map and a distance dictionary as a two-item
return value. A mapping skill that computes both but returns only the map
matches the topic but violates the requested return interface. For a
second query asking for the one-item interface, the same skill becomes
the compatible choice. Compatibility thus depends on the query, even
when the skills share a capability. A retrieved list can contain both
the compatible skill and its query-inappropriate sibling.

We formalize this problem as \textbf{\emph{same-capability risk-exposure retrieval}}.
For each query, we identify one locally helpful skill and a marked sibling
with a specific conflicting span. Three complementary metrics distinguish
helpful retrieval from conflicting exposure. Recall@K measures whether the helpful
skill appears in the top-$K$ list; harmful sibling rate (HSR@K) measures
whether the marked sibling appears. CleanHit@K combines the two conditions:
the list contains the helpful skill and excludes that sibling. Together,
they separate missed helpful skills, helpful hits accompanied by a conflict,
and helpful hits without that conflict
(Figure~\ref{fig:skillresolve-concept}\label{mention:fig:skillresolve-concept}).

\textbf{SameCapRisk-Bench} contains 890 units and 1,314 queries under five
mechanisms and 12 conflict types. Each record links a query requirement to
a conflicting skill passage and a supporting source or local check.
There are 466 single-query contracts,
387 source-role pairs (774 queries), and 37 author-derived output-interface
pairs (74 queries). We reserve 55 single-query units and 12 interface pairs
for development: 67 units and 79 queries in total. This leaves 823 units
and 1,235 queries for the held-out main evaluation. Source-role queries
use their own candidate pool; the other two protocols share a public-skill
pool. Because the protocols differ in task source, construction, and pool,
we report their scores separately. Conflict-type results and sample counts
further show which requirements account for the pooled performance.

Four released skill retrievers achieve pooled Recall@3 of 0.903--0.944,
yet their HSR@3 remains
0.344--0.393. Exposure is much higher on source-task contracts
(HSR@3 of 0.737--0.881) than on controlled source-role queries
(0.099--0.12). Even on queries where all seven neural methods retrieve
both marked skills in their top 20, they frequently rank the conflicting
skill above the helpful one on source-task queries, while source-role
ordering is nearly saturated. Reranking identical candidates also yields
much larger CleanHit gains on source-role queries than on source-task
contracts.

Reranking also depends on what the model can see. Truncation can remove
the passage where a skill's contract differs from its sibling's, leaving
identical inputs for two different skills. With the same BGE top-20 candidates, increasing the
reranker's budget from 512 to 4,096 tokens lowers source-task HSR@1 from
0.319 to 0.236. CleanHit@3 rises only from 0.056 to 0.063; the pooled gain's
95\% interval spans zero. Longer inputs improve which skill ranks first,
while conflicting siblings largely remain in the returned list. This
distinction motivates separate checks of the model's input, the relative
order of the two skills, and the composition of the final list.
Our contributions are:
\begin{itemize}
    \item A formulation of same-capability risk-exposure retrieval:
    skills share a capability, but a specific contract conflict makes one
    incompatible with the current query.
    \item A benchmark of 890 units linking query requirements to skill
    contrasts and supporting evidence, organized by a two-level taxonomy
    and three query protocols, with checks for candidate-side cues and
    dependence on the query. Resource:
    \url{https://github.com/jdding/SameCapRisk}.
    \item An evaluation design combining Recall, HSR, and CleanHit with
    common-candidate ordering and matched reranking and input-budget
    comparisons, to distinguish recovery, ordering, and list-composition
    errors.
    \item An empirical analysis showing frequent conflict exposure despite
    high helpful recall, protocol-dependent reranking gains, and improved
    first-place choice from longer inputs with little change in clean
    helpful retrieval.
\end{itemize}

%% file: sections/2_related_work.tex
\section{Related Work}

Existing benchmarks evaluate broad retrieval quality or downstream utility:
BEIR spans heterogeneous retrieval tasks \citep{beir2021}, WebArena measures
interactive Web task completion \citep{webarena2023}, and ToolLLM evaluates
API selection and use \citep{toollm2023}. Skill-use benchmarks extend this
evaluation to execution gains, repository-based utility, and bottlenecks
between retrieval and execution
\citep{skillsbench2026,sweskillsbench2026,sra2026}.
Within this pipeline, tool-retrieval methods learn relevance or improve
candidate selection through hierarchy-aware and iterative reranking
\citep{autotools2025,toolret2025,toolrerank2024,toolqp2026}.
Skill retrievers differ in the information used to distinguish candidates:
SkillRouter uses full skill bodies, while SkillSight removes shared
descriptive background \citep{skillrouter2026,skillsight2026}; other methods
revise overlapping descriptions, organize skill groups and dependencies,
or adapt ranking to execution state
\citep{singlerewrite2026,goskills2026,graphofskills2026,skillreranker2026}.
SkillRet credits useful or equivalent alternatives in public-library
retrieval, whereas R3-Skill learns query-conditioned compatibility of a
retrieved set from rejected combinations
\citep{skillret2026,skillisnotdocument2026}.
Our evaluation asks a narrower question: does retrieval expose a
same-capability alternative whose recorded passage conflicts with the
current requirement, even when the helpful skill is present? Paired
queries test this distinction by changing the requirement while keeping
both candidate skills fixed.

NLP evaluation has long used hard distractors to expose failures hidden by
surface relevance. Adversarial reading-comprehension examples add misleading
sentences that preserve the human answer \citep{jia2017adversarialqa},
and contrast sets perturb examples to probe local
decision boundaries \citep{gardner2020contrastsets}.
\rev{SameCapRisk-Bench applies this local-comparison logic to a query and two
skills with the same broad capability. A contrast is tied to a specific
resource, procedure, precondition, output, or evidence requirement. In the
paired protocols, changing the requirement exchanges the roles of the fixed
skills.}

Broader agent-risk benchmarks evaluate unsafe tool use and prompt injection at
execution time \citep{toolemu2024,agentdojo2024}.
Skill-specific benchmarks examine harmful-skill misuse, runtime trust
failures, and skill-facing attack surfaces
\citep{harmfulskillbench2026,agenttrap2026,skillsafetybench2026}.
MalSkillBench adds runtime-verified malicious-skill ground truth spanning
instruction and code behavior, while SkillGuard treats skills as
permission-bearing executable artifacts governed through manifests, runtime
access control, and audit signals \citep{malskillbench2026,skillguard2026}.
These studies evaluate malicious behavior or unsafe outcomes during skill
execution. We instead measure exposure at retrieval time: a returned skill
conflicts with the current requirement but may fit another query.
The label does not assert malicious intent or an observed execution failure.

RRF combines lexical rankings \citep{cormack2009rrf}; BGE-M3 provides dense
retrieval \citep{bge_m3_2024}. Cross-encoders score query--skill
pairs, and prompted listwise rerankers compare candidates jointly
\citep{rankgpt2023}. Our general baselines are RRF and BGE-M3. We also
rerank the same BGE top-20 candidates with a BGE cross-encoder or Qwen2.5-7B
\citep{qwen25technicalreport2025}. This matched comparison separates
candidate recovery from subsequent ordering. A separate control keeps
retrieval scores fixed and compares
selection from text-based clusters with selection from the benchmark's
released skill-pair relations.

%% file: sections/3_problem_method.tex
\section{Task, Benchmark, and Metrics}

\begin{table}[t]
\caption{Benchmark composition by source. Main counts exclude 67 development
units (79 queries). The public and source-role candidate pools contain 7,615
and 774 candidates, respectively.}
\label{tab:benchmark-composition}
\centering
\small
\setlength{\tabcolsep}{3.0pt}
\begin{tabular*}{\linewidth}{@{}l@{\extracolsep{\fill}}rrrr@{}}
\toprule
& \multicolumn{2}{c}{Full benchmark} & \multicolumn{2}{c}{Held-out} \\
\cmidrule(lr){2-3}\cmidrule(l){4-5}
Source & Units & Queries & Units & Queries \\
\midrule
\input{generated/working890_source_rows.tex} 
\bottomrule
\end{tabular*}
\end{table}

\subsection{Failure Mode and Benchmark Unit}

The basic data item is a \emph{benchmark unit}: two same-capability skills
paired with one or two queries. A retriever
may reach the correct capability family while exposing a sibling whose
contract conflicts with the current query. Formally,
\[
\cD=\{(\query_i,\cC_i,\spos_i,\sneg_i,g^\star,e_i)\}_{i=1}^{N},
\]
where $N$ counts query cases, $\query_i$ is a query, $\cC_i$ its fixed candidate pool, $\spos_i$ an
admitted helpful skill, $\sneg_i$ a query-specific conflicting sibling, $g^\star$ the
released family relation, and $e_i$ the admission evidence. The two marked
skills share a capability family but differ in an operational contract or in
what their evidence can establish. The label is relational: helpful means
that the relevant skill passage meets the tested query requirement.
Admission applies to this local contract.

\subsection{Benchmark Construction}
\label{sec:construction}

Figure~\ref{fig:benchmark-pipeline}\label{mention:fig:benchmark-pipeline}
summarizes construction: we form query--skill contrasts, check their local
contracts, and freeze the inputs for retrieval evaluation. SameCapRisk-Bench
uses three query protocols over two candidate pools.
\rev{\emph{Single-query contracts} contain 466 units derived from source tasks,
each with one query and a skill that conflicts with a concrete requirement.}
\emph{Source-role pairs}
contain 387 controlled units and 774 queries: two well-formed skills exchange
helpful and conflicting roles when the requested evidence role changes.
\emph{Output-interface pairs} contain 37 author-derived units and 74 queries:
two queries impose different observable output requirements, including
returns, exceptions, or side effects, and reverse which of the same two
skills fits. Source-role queries
rank against a 774-candidate pool; the other two
protocols use a separate 7,615-candidate public pool.

Overall there are 890 units and 1,314 query cases. The main evaluation
contains 411 single-query, 774 source-role, and 50 output-interface cases,
or 823 units and 1,235 queries. The remaining 67 units (79 queries) are
development-only. Paired queries and source-query lineages stay within
one split.
Table~\ref{tab:benchmark-composition}\label{mention:tab:benchmark-composition}
reports the source composition and split sizes.

Single-query units test whether retrieval preserves a source task's
requirement among public-library candidates. In source-role pairs, a query
requests a particular kind of evidence, such as a decision rule rather
than a case record. In output-interface pairs, it specifies the required
result or side effect for the same broad task.
The paired protocols therefore test whether selection changes with the
requirement, rather than consistently favoring one of the fixed skills.

Public collections supply the task requirements and skill workflows from
which we construct contrasts. \rev{For each single-query unit, we identify a concrete
requirement in the full source query: an input condition, a resource or
parameter, an operation, or an observable output. The helpful skill follows
that requirement in its workflow. We change the corresponding part of a
same-capability sibling while keeping the rest of the capability comparable,
then check both skills against the same requirement and evidence.}
Admission requires a specific conflicting operation or passage.
Numerical and logical tasks specify the computation or preconditions to
check; software and skill workflows specify fields, resources, schemas,
or exception behavior. For output-interface pairs, we derive the two
queries from explicit requirements on observable behavior.

We organize conflicts into five primary mechanisms, assigning each unit
one mechanism and one type. \textbf{Resource binding} concerns
\emph{which object or field} a workflow uses; \textbf{Procedure} concerns
\emph{which computation or branch} it follows; \textbf{Applicability}
concerns \emph{when} it is valid; \textbf{Output behavior} concerns
\emph{what it returns or changes}; \textbf{Evidence role} concerns
\emph{what the retrieved material can establish}.
We call the first four \emph{operational contracts} because they govern
what a skill reads, does, accepts, or returns; \emph{evidence role} instead
governs what a retrieved artifact can justify. The mechanisms expand into 12
fixed types, including resource identity, field/parameter binding, computation
rule, control-flow selection, precondition mismatch, representation schema,
return/exception/side effect, and five evidence-role distinctions. The
smallest single correction that restores the query-required
contract determines the primary label; secondary effects remain provenance
notes. Appendix~\ref{app:taxonomy} gives the boundary rules.
All 12 types contain at least 32 units overall. Some types have fewer units
in the main split, so we report sample counts with type-level comparisons.

\begin{table}[t]
\caption{Full and main-evaluation coverage by conflict type. Bold italic rows
denote primary mechanisms; indented rows denote their fixed subtypes.
Paired units contribute two query cases.}
\label{tab:taxonomy-coverage}
\centering
\small
\setlength{\tabcolsep}{2.3pt}
\begin{tabular*}{\linewidth}{@{}l@{\extracolsep{\fill}}rrrr@{}}
\toprule
& \multicolumn{2}{c}{Full benchmark} & \multicolumn{2}{c}{Main evaluation} \\
\cmidrule(lr){2-3}\cmidrule(l){4-5}
Conflict type & Units & Queries & Units & Queries \\
\midrule
\input{generated/working890_taxonomy_rows.tex} 
\bottomrule
\end{tabular*}
\end{table}
In a Code Review Triage unit, both skills address code review, but their
evidence serves different purposes. The \emph{Policy Rubric Procedure}
interprets merge-request criteria and decision thresholds; the
\emph{Case Record Procedure} summarizes facts and outcomes from a concrete
change record. A query asking for criteria to guide an upcoming decision
requires the rubric. A query asking what happened in an already handled
case requires the case record. These queries exchange the helpful and
conflicting roles while both skill texts stay fixed, including their shared
instructions to name the evidence, explain its interpretation, and preserve
caveats. The contrast is \emph{Evidence role: Normative vs. empirical}
(Table~\ref{tab:taxonomy-coverage}\label{mention:tab:taxonomy-coverage}): a decision rule does not
establish the facts of a particular event, and a case record does not
supply the rule for a new decision.

For the family-based selection control, we connect the two skills in each
admitted pair and define a family as a connected component of these edges.
A background skill outside any admitted pair forms a family containing only
itself. Thus families encode the released pair relationships, not semantic
clusters of the library. Public text-only retrievers receive candidate text
without these edges.

Units used for benchmark development retain that assignment; they do not
enter the held-out evaluation. All queries derived from one source query,
and both queries of a paired unit, stay in the same split. We check connected
components of source lineage and identical texts for split crossings.
Source-role pairs are evaluation-only. These assignments were fixed before
the retrieval runs.

\subsection{Benchmark Quality}
\label{sec:construction-checks}

A single-query unit is admitted only when the source query states the
requirement, the helpful skill's relevant passage satisfies it, and the
sibling's passage contradicts it. The cited source, checker, or oracle
evidence must support that comparison for the recorded query and skill
texts. Paired units require two well-formed skills, two queries that exchange
their roles, and the same local judgment rule on both sides.

\begin{table}[!b]
\caption{Held-out retrieval on 1,235 queries. Each cell shows the point
estimate and a 95\% group-resample interval. Bold marks the best point estimate.}
\label{tab:main}
\centering
\footnotesize
\setlength{\tabcolsep}{1.0pt}
\begin{tabular*}{\linewidth}{@{}l@{\extracolsep{\fill}}ccc@{}}
\toprule
Method & Recall@3 $\uparrow$ & HSR@3 $\downarrow$ & CleanHit@3 $\uparrow$ \\
\midrule
\input{generated/working890_main_rows.tex} 
\bottomrule
\end{tabular*}
\end{table}

\begin{table*}[t]
\caption{Retrieval within each held-out protocol at $K=3$. Single-query and
output-interface cases share the public pool; source-role cases use their
own pool. Each method is evaluated on the same queries within a protocol.}
\label{tab:protocol}
\centering
\small
\setlength{\tabcolsep}{2pt}
\begin{tabular*}{\textwidth}{@{}l@{\extracolsep{\fill}}rrrrrrrrr@{}}
\toprule
& \multicolumn{3}{c}{Single-query ($n=411$)}
& \multicolumn{3}{c}{Source-role ($n=774$)}
& \multicolumn{3}{c}{Output-interface ($n=50$)} \\
\cmidrule(lr){2-4}\cmidrule(lr){5-7}\cmidrule(l){8-10}
Method & Recall@3 $\uparrow$ & HSR@3 $\downarrow$ & CleanHit@3 $\uparrow$
& Recall@3 $\uparrow$ & HSR@3 $\downarrow$ & CleanHit@3 $\uparrow$
& Recall@3 $\uparrow$ & HSR@3 $\downarrow$ & CleanHit@3 $\uparrow$ \\
\midrule
\input{generated/working890_protocol_rows.tex} 
\bottomrule
\end{tabular*}
\end{table*}

The check depends on the contract: mathematical or algorithmic certificates
check numerical parameters and preconditions; interface checks compare the
observable behavior with the query's exact requirement; evidence-role checks
compare what each procedure can establish with the requested kind of
evidence. Supporting records include 160 local certificates, 37 tool/backend
or numerical outcome contrasts, and 387 static source-role comparisons.
The remaining records check requested interfaces, compare skill operations
with source requirements, or identify changed code and its source.
These are checks of the stated local contrast, not certificates of a
skill's entire execution. Appendix~\ref{app:admission} details their coverage
and explains how evidence remains tied to the assessed texts.

We also evaluated 96 stratified held-out units (128 queries) using the
DeepSeek-V4-Pro and GLM-5.3 APIs. Each model received the complete query
and both skill texts with labels hidden, once in each candidate order.
Both models matched the recorded label
for every query in both orders on 88/96 units. Reviewing the other 8
units identified 2 query--skill mismatches; a targeted same-class scan of
the 890-unit roster found 3 more. We corrected these 5 units and
recomputed the retrieval results. On original-order queries, the models agreed on
124/128 judgments ($\kappa=0.94$). Appendix~\ref{app:cross-family-audit}
details the sample, position checks, and construction fixes.

For retrieval, inference files expose only neutral identifiers, query text, and candidate
text. Helpful/risky roles, taxonomy labels, evidence, source paths, split
roles, and family relations are stored separately. Shared rules were used to
remove known asymmetric wording before data freeze.

To test whether candidate wording reveals the label without the query,
we fit a character
TF--IDF classifier on training candidates in each of five group-disjoint
folds, keeping both candidates and all units in a group together.
It reaches 59.66\% on the 466 single-query units, rising to
73.44\% for control-flow and 72.73\% for representation/schema contrasts.
Residual candidate-side predictability is therefore concentrated unevenly
across conflict types.

To test choice with the query present, a fixed Qwen3-4B probe receives the
complete query and both skill texts
under neutral A/B identifiers for a direct pairwise choice. It uses greedy
decoding and rejects inputs exceeding its context limit.
It selects the compatible side on 298/466
single-query cases, 765/774 source-role queries, and 65/74 output-interface
queries. The remaining 168 single-query responses comprise 77 choices of
the alternative candidate, 28 abstentions, and 63 format-invalid outputs.
The probe is nearly perfect on source-role matching, whereas
single-query contracts remain harder. The probe selects the correct skill in
both original and mirrored candidate orders for 43/64 selected tests. Across 2,493 probe outputs,
148 format-invalid responses remain in the denominator. In the query-mismatch
test, we change the query but keep both skills fixed, then measure how
often the probe still selects the original query's helpful skill. This
measures dependence on the query, not accuracy on the replacement task.
Appendix~\ref{app:construct-validity} gives the
model revision, decoding settings, and perturbation protocol. These checks
describe construction and pairwise choice; the
full-pool retrieval experiments below ask a different question.
Appendix~\ref{app:probe-outcomes} separates valid alternative choices from
abstentions and malformed responses.

\subsection{Retrieval Metrics}

We evaluate whether retrieval supplies a skill that meets the query's
requirement without also exposing its conflicting sibling. For the $n>0$
held-out query cases, a system returns an ordered list
$R_K(\query_i)$ of at most $K$ distinct candidates from $\cC_i$.
With $\mathbb I$ denoting an indicator, Recall@K measures whether the
helpful skill is retrieved:
\begin{equation}\label{eq:recall}
\mathrm{Recall}@K=\frac{1}{n}\sum_{i=1}^{n}
\mathbb I[\spos_i\in R_K(\query_i)].
\end{equation}
Harmful sibling rate (HSR@K)
measures exposure of the marked risky sibling:
\begin{equation}\label{eq:hsr}
\mathrm{HSR}@K = \frac{1}{n}\sum_{i=1}^{n}
\mathbb{I}[\sneg_i\in R_K(\query_i)].
\end{equation}
A relevance-only scorer can retrieve both siblings, yielding high Recall
and high HSR together. CleanHit@K captures the joint outcome:
\begin{equation}\label{eq:cleanhit}
\mathrm{CleanHit}@K = \frac{1}{n}\sum_{i=1}^{n}
\mathbb{I}[\spos_i\in R_K(\query_i)\wedge\sneg_i\notin R_K(\query_i)].
\end{equation}
We use these metrics at $K\in\{1,3,5,20\}$ to distinguish helpful recovery,
marked conflict exposure, and helpful retrieval without that exposure.
HSR concerns the annotated sibling in a fixed library, not every possible
incompatible candidate. CleanHit also differs from pairwise ordering: a list with
the helpful skill first and the sibling second has correct relative order
but CleanHit@3 of zero. We therefore examine both list composition and
relative order. These three metrics measure list membership; relative rank
and model-input visibility require separate checks. At $K=1$, HSR is the
marked-sibling first-place rate and CleanHit equals Recall.
Table~\ref{tab:main} reports these metrics over the held-out queries;
Table~\ref{tab:protocol} separates the three query protocols.

%% file: generated/working890_source_rows.tex
BigCodeBench~\citep{bigcodebench2025} & 113 & 147 & 46 & 68 \\
CHAMP~\citep{champ2024} & 59 & 59 & 59 & 59 \\
LogicBench~\citep{logicbench2024} & 15 & 15 & 15 & 15 \\
MedCalcBench~\citep{medcalcbench2024} & 38 & 38 & 38 & 38 \\
SkillLearnBench~\citep{skilllearnbench2026} & 42 & 43 & 42 & 43 \\
SkillsBench~\citep{skillsbench2026} & 67 & 69 & 67 & 69 \\
TheoremQA~\citep{theoremqa2023} & 159 & 159 & 159 & 159 \\
ToolQA~\citep{toolqa2023} & 10 & 10 & 10 & 10 \\
\midrule
Source-role pairs (ours) & 387 & 774 & 387 & 774 \\
\midrule
Total & 890 & 1,314 & 823 & 1,235 \\

%% file: generated/working890_taxonomy_rows.tex
\multicolumn{5}{@{}l}{\textbf{\emph{Resource binding}}} \\
\quad Resource identity & 72 & 72 & 65 & 65 \\
\quad Field / parameter binding & 124 & 124 & 118 & 118 \\
\multicolumn{5}{@{}l}{\textbf{\emph{Procedure}}} \\
\quad Computation rule & 126 & 126 & 118 & 118 \\
\quad Control-flow selection & 32 & 32 & 17 & 17 \\
\multicolumn{5}{@{}l}{\textbf{\emph{Applicability}}} \\
\quad Precondition mismatch & 57 & 57 & 54 & 54 \\
\multicolumn{5}{@{}l}{\textbf{\emph{Output behavior}}} \\
\quad Representation schema & 57 & 59 & 41 & 43 \\
\quad Return / exception / side effect & 37 & 74 & 25 & 50 \\
\multicolumn{5}{@{}l}{\textbf{\emph{Evidence role}}} \\
\quad Aggregate vs. instance & 59 & 118 & 59 & 118 \\
\quad Intent vs. outcome & 98 & 196 & 98 & 196 \\
\quad Measurement vs. calibration & 37 & 74 & 37 & 74 \\
\quad Normative vs. empirical & 138 & 276 & 138 & 276 \\
\quad State change vs. rationale & 53 & 106 & 53 & 106 \\

%% file: generated/working890_main_rows.tex
RRF & 0.781\,[0.727,0.837] & 0.358\,[0.278,0.440] & 0.440\,[0.367,0.525] \\
BGE-M3 & 0.811\,[0.701,0.884] & 0.329\,[0.286,0.380] & 0.495\,[0.397,0.567] \\
\midrule
SkillRet & 0.903\,[0.870,0.929] & 0.344\,[0.294,0.399] & 0.564\,[0.506,0.624] \\
SkillSight & 0.926\,[0.880,0.959] & 0.388\,[0.352,0.427] & 0.547\,[0.474,0.606] \\
SkillRouter & 0.939\,[0.923,0.955] & 0.386\,[0.344,0.430] & 0.564\,[0.515,0.610] \\
R3-Skill & \textbf{0.944}\,[0.921,0.965] & 0.393\,[0.340,0.453] & 0.563\,[0.509,0.613] \\
\midrule
BGE reranker & 0.863\,[0.826,0.893] & \textbf{0.297}\,[0.258,0.339] & \textbf{0.580}\,[0.516,0.634] \\
Qwen2.5-7B & 0.849\,[0.789,0.893] & 0.324\,[0.287,0.364] & 0.550\,[0.491,0.603] \\

%% file: generated/working890_protocol_rows.tex
RRF & 0.727 & 0.669 & 0.092 & 0.848 & 0.202 & 0.654 & 0.200 & 0.220 & 0.000 \\
BGE-M3 & 0.686 & 0.667 & 0.051 & 0.873 & 0.114 & 0.762 & 0.880 & 0.880 & 0.000 \\
\midrule
SkillRet & 0.766 & 0.737 & 0.046 & 0.977 & 0.102 & 0.875 & 0.880 & 0.860 & 0.020 \\
SkillSight & 0.876 & 0.827 & 0.075 & 0.953 & 0.120 & 0.833 & 0.920 & 0.920 & 0.000 \\
SkillRouter & 0.847 & 0.852 & 0.029 & 0.988 & 0.103 & 0.885 & 0.940 & 0.940 & 0.000 \\
R3-Skill & 0.893 & 0.881 & 0.044 & 0.973 & 0.099 & 0.875 & 0.920 & 0.920 & 0.000 \\
\midrule
BGE reranker & 0.720 & 0.696 & 0.056 & 0.963 & 0.071 & 0.893 & 0.500 & 0.520 & 0.040 \\
Qwen2.5-7B & 0.696 & 0.681 & 0.085 & 0.925 & 0.115 & 0.813 & 0.920 & 0.620 & 0.300 \\

%% file: sections/4_experiments_working890.tex
\section{Benchmark Evaluation}
\label{sec:evaluation}

\begin{table*}[!t]
\caption{Held-out CleanHit@3 by conflict type (higher is better).
Green shading uses a shared 0--1 scale; bold marks the row-best point
estimate, including ties. Q, U, and G count queries, units, and groups.
BGE-R and Qwen rerank the same BGE top-20 candidates. Rows with few
groups support descriptive comparisons.}
\label{tab:subtype-results}
\centering
\small
\definecolor{cleanhitgreen}{HTML}{78B99A}
{\setlength{\fboxsep}{2pt}
CleanHit@3 $\uparrow$\quad
\colorbox{cleanhitgreen!0}{\strut 0.00}\,
\colorbox{cleanhitgreen!25}{\strut 0.25}\,
\colorbox{cleanhitgreen!50}{\strut 0.50}\,
\colorbox{cleanhitgreen!75}{\strut 0.75}\,
\colorbox{cleanhitgreen!100}{\strut 1.00}\par}
\smallskip
\setlength{\tabcolsep}{2pt}
\begin{tabular*}{\textwidth}{@{}l@{\extracolsep{\fill}}rrrrrrrrr@{\hspace{7pt}}rr@{}}
\toprule
Conflict type & Q & U & G & RRF & BGE & SkillRet & SkillSight & SkillRouter & R3 & BGE-R & Qwen \\
\midrule
\input{generated/working890_subtype_rows.tex} 
\bottomrule
\end{tabular*}
\end{table*}

\subsection{Protocol and Methods}
\label{sec:retrieval-protocol}

The held-out evaluation contains 1,235 queries from 823 units. The
single-query and output-interface protocols use the 7,615-candidate public
pool; source-role queries use a separate 774-candidate pool. Methods rank
within the relevant pool. We report query-micro Recall, marked-sibling
exposure (HSR), and CleanHit at $K=3$, representing a short list in which
multiple skills can enter the agent's context. We test other list sizes at
$K=1,5,20$ and alternative weights in Appendix~\ref{app:retrieval-analyses}.
All eight methods operate on query and candidate text, with held-out
labels, taxonomy and family relations kept separate. Grouping controls are
reported separately.

RRF combines word and character lexical ranks~\citep{cormack2009rrf};
BGE-M3 supplies a general dense baseline. SkillRet, SkillSight, SkillRouter,
and R3-Skill use released skill-retrieval methods and checkpoints
\citep{skillret2026,skillsight2026,skillrouter2026,skillisnotdocument2026}.
The R3-Skill checkpoint is pinned for this evaluation. BGE reranking and
prompted Qwen2.5-7B listwise reranking both reorder the same BGE top-20
candidates
\citep{bge_m3_2024,qwen25technicalreport2025,rankgpt2023}.
The main BGE reranker uses a combined query--document budget of 512 tokens;
Section~\ref{sec:interventions} tests 4,096 tokens separately.
Appendix~\ref{app:retrieval-analyses} specifies listwise output completion;
Appendix~\ref{app:probe-outcomes} separately reports probe response validity.
The 6,660 background skills originate from SkillRet's public corpus, making
its released checkpoint an in-domain comparator. Available training records
leave exact document-level overlap for other public checkpoints unresolved.

For uncertainty, we resample construction groups within each protocol,
then combine protocol estimates at their fixed query
weights (5,000 shared draws). The intervals are conditional on this
mixture and grouping; the source-role protocol has ten such groups.

\subsection{Helpful Retrieval and Marked Exposure}

We first ask whether methods that retrieve the helpful skill also keep its
conflicting sibling out of the returned list (Table~\ref{tab:main}). For the four specialized skill
retrievers, Recall@3 is 0.903--0.944, but HSR@3 is 0.344--0.393 and
CleanHit@3 only 0.547--0.564. High helpful recall therefore does not imply
clean selection. R3-Skill has the highest Recall point estimate, whereas
BGE reranking has the highest CleanHit point estimate despite lower Recall.
The choice of metric changes which outcome appears best; these point
estimates do not by themselves establish a robust ordering of methods.

Table~\ref{tab:protocol} shows why the pooled score needs a protocol-level
reading. On 411 single-query contracts, specialized retrievers expose the
marked sibling in 0.737--0.881 of top-3 lists, leaving CleanHit@3 at
0.029--0.075. On 774 source-role queries, HSR@3 is 0.099--0.12 and
CleanHit@3 is 0.833--0.885. The 50 output-interface queries again show
high exposure (HSR@3 of 0.86--0.94), despite reversing which interface
the query requires. Thus the high pooled CleanHit values largely reflect
the 62.7\% source-role share, not uniformly clean lists across protocols.
In a post-hoc check on the fixed rankings, removing those 774 queries
makes Qwen listwise the CleanHit leader on the remaining 461 (0.108,
versus 0.054 for BGE reranking). These protocols also differ in source,
pool, construction, and query length; the contrast is not an isolated
effect of conflict type.

Exposure reaches different positions in those lists. For the four
specialized retrievers, single-query HSR@1 is 0.248--0.292, whereas
source-role HSR@1 is 0--0.003. Source-role conflicts therefore appear
mainly below first place; on source-task contracts, both the first choice
and the rest of the list remain problematic. This is why first-place
accuracy alone would miss part of the measured exposure.

Methods also trade helpful recovery against clean selection. On
single-query contracts, RRF has higher CleanHit@3 than the neural-method
point estimates (0.092), but lower Recall@3 (0.727 versus 0.766--0.893
for the four specialized retrievers). On output-interface queries, RRF
finds the helpful skill in only 10/50 top-3 lists and 32/50 top-20 lists;
every top-3 helpful hit includes the sibling. Qwen instead reaches
CleanHit@3 of 0.3 there, versus at most 0.04 for other neural methods.
That advantage concerns 25 author-derived exact-interface pairs, not a
general claim about other operational contracts.

\subsection{Where Contract-Level Errors Occur}

The conflict-type results in Table~\ref{tab:subtype-results} show that a
high protocol average can conceal different failures. Resource binding
and procedure cover 318 main queries, yet the four specialized retrievers
have CleanHit@3 of only 0.022--0.060 and 0.022--0.052, respectively.
The 54 applicability queries also show frequent marked exposure
(HSR@3 of 0.759--1). Evidence role, with 770 queries from 385 units
and nine groups, ranges from CleanHit@3 of 0.551--0.737 on
aggregate-versus-instance contrasts to 0.959--1 on
measurement-versus-calibration contrasts. These values show variation
within evidence role, not just between protocol averages. Evidence role
nearly coincides with the source-role protocol, while output behavior
mixes single-query and output-interface cases. Type and construction
conditions therefore cannot be separated by this table alone.
Table~\ref{tab:taxonomy-coverage} gives full and main-split supports;
comparisons with 17 control-flow or 25 return/exception main units are
descriptive.

Within single-query contracts, the taxonomy separates two reasons for
low CleanHit. On 118 field/parameter queries, the specialized retrievers
achieve Recall@3 of 0.864--0.949 but CleanHit@3 of 0.017--0.068: they
often find a suitable skill without excluding the sibling. On 118
computation-rule queries, Recall@3 is lower (0.695--0.873) and
CleanHit@3 remains low (0.025--0.059), indicating both recovery and
list-composition errors. On the 54 precondition queries, every
SkillRouter top-3 list contains both marked skills. The useful distinction
is whether the helpful skill is missing or is present alongside its
conflicting alternative.

We next ask whether the methods choose the right representative once
both skills are available. All seven neural methods retrieve both marked
candidates in their top 20 on the same 543 main queries: 309 single-query,
188 source-role, and 46 output-interface cases. On the 309 single-query
cases, the helpful skill ranks above its sibling on only 0.521--0.696
of queries, compared with 0.952--1 on the 188 source-role cases.
Recovery alone therefore does not resolve source-task ordering, and a
correctly ordered pair can still leave both skills in the top 3. This
conditional ordering check supplements, rather than replaces, the
full-pool scores over all 1,235 queries. The next analysis examines
whether changing model inputs and grouping changes these outcomes.

\subsection{Reranking, Input Length, and Family Information}
\label{sec:interventions}

\begin{figure}[t]
\centering
\includegraphics[width=\columnwidth]{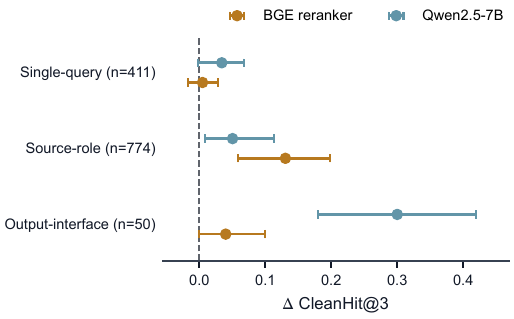}
\caption{CleanHit@3 changes relative to BGE-M3 after reranking its top-20 candidates
with the 512-token BGE reranker or Qwen listwise.
Points are query-matched differences; bars are 95\% group-resample intervals
within each protocol.}
\Description{Three horizontal rows show reranking changes for single-query,
source-role, and output-interface protocols. BGE's single-query interval
crosses zero, while its larger source-role gain is positive.}
\label{fig:matched-reranking}
\end{figure}

\paragraph{Same-candidate reranking.}
To test ordering without changing candidate recovery, we rerank identical
BGE top-20 sets (Figure~\ref{fig:matched-reranking}). BGE reranking raises
CleanHit@3 by 0.085 over BGE dense ranking on the 1,235 main queries
(140 query-level gains, 35 losses; paired group-resample 95\% interval
[0.04, 0.128]). Yet its gain comes mainly from source-role queries
(+0.13 [0.059, 0.199]); the single-query change is only +0.005
[-0.017, 0.028]. Qwen listwise reranking gains +0.055 [0.024, 0.096]
pooled and +0.034 [-0.002, 0.067] on single-query contracts, with its
largest point gain on the 50 author-derived output-interface queries.
Thus reranking the same candidates can improve CleanHit, but a pooled
gain need not extend to source-task contracts with high marked exposure.

\paragraph{\rev{Longer inputs for dense retrieval.}}
\rev{We rerun the BGE dense encoder with an input budget of 4,096 rather
than 512 tokens. This changes 447 of 1,235 top-20 sets. At $K=3$ it adds
27 helpful hits and 31 marked exposures, with 14 CleanHit gains and 15
losses. All 774 source-role queries retain the same ordered top-20 lists;
changes occur in the two public-pool protocols. Longer inputs recover
additional helpful skills but also expose additional siblings, leaving
the overall CleanHit score nearly unchanged (0.495 to 0.494).}

\paragraph{Input identity and reranking.}
\rev{We separately reconstruct the text delivered to the BGE reranker and
check whether the two marked candidates remain distinguishable. With the recorded
512-token reranker input, the marked candidates become token-identical for
308/1,235 main queries, including 302/411 single-query contracts. Reconstructing
the inputs with a 4,096-token budget reduces these counts to 61 and 57.
The Qwen listwise excerpt
likewise leaves 314/1,235 marked pairs text-identical.}

Increasing the same BGE reranker's budget from 512 to 4,096 tokens, with
BGE's top-20 candidate sets fixed, lowers single-query HSR@1 by
8.27~pp (0.319 to 0.236). CleanHit@3 changes from 0.056 to 0.063.
Source-role rankings are unchanged: all their reranker inputs already fit
within 512 tokens. On output-interface queries, CleanHit@3 changes from
0.04 to 0.02.
The pooled CleanHit gain is $+0.0016$ (16 gains, 14 losses; paired
95\% group-resample interval [$-0.0065$, $0.01$]). Among the 247 main queries whose marked candidates
change from token-identical at 512 tokens to distinct at 4,096, CleanHit@3
improves on nine queries and worsens on eight. The larger budget reduces
first-place sibling errors, while its effect on CleanHit@3 remains small
and uncertain. The larger budget also changes text beyond the decisive
contract span, so a span-level check is needed to isolate that mechanism.

\paragraph{Grouping information.}
We ask how the grouping used to select representatives changes the
returned list when retrieval scores stay fixed. With SkillSight's score
order, public text clusters yield Recall@3, HSR@3, and CleanHit@3 of
0.834, 0.155, and 0.778. Using the benchmark's released pair families
changes these to 0.842, 0.105, and 0.842: marked exposure falls while
helpful recall stays similar. This control shows sensitivity to grouping
information, not a deployable gain. It uses family edges withheld from
public methods and therefore belongs outside their leaderboard; using
families in a new library would first require constructing them from
available information.

%% file: generated/working890_subtype_rows.tex
\multicolumn{12}{@{}l}{\textbf{\emph{Resource binding}}} \\
Resource identity & 65 & 65 & 56 & \cellcolor{cleanhitgreen!15.4}\textbf{0.154} & \cellcolor{cleanhitgreen!1.5}0.015 & \cellcolor{cleanhitgreen!4.6}0.046 & \cellcolor{cleanhitgreen!4.6}0.046 & \cellcolor{cleanhitgreen!6.2}0.062 & \cellcolor{cleanhitgreen!1.5}0.015 & \cellcolor{cleanhitgreen!0.0}0.000 & \cellcolor{cleanhitgreen!7.7}0.077 \\
Field / parameter binding & 118 & 118 & 117 & \cellcolor{cleanhitgreen!6.8}\textbf{0.068} & \cellcolor{cleanhitgreen!2.5}0.025 & \cellcolor{cleanhitgreen!5.1}0.051 & \cellcolor{cleanhitgreen!6.8}\textbf{0.068} & \cellcolor{cleanhitgreen!1.7}0.017 & \cellcolor{cleanhitgreen!2.5}0.025 & \cellcolor{cleanhitgreen!3.4}0.034 & \cellcolor{cleanhitgreen!5.9}0.059 \\
\multicolumn{12}{@{}l}{\textbf{\emph{Procedure}}} \\
Computation rule & 118 & 118 & 117 & \cellcolor{cleanhitgreen!3.4}0.034 & \cellcolor{cleanhitgreen!6.8}\textbf{0.068} & \cellcolor{cleanhitgreen!2.5}0.025 & \cellcolor{cleanhitgreen!5.9}0.059 & \cellcolor{cleanhitgreen!2.5}0.025 & \cellcolor{cleanhitgreen!5.9}0.059 & \cellcolor{cleanhitgreen!2.5}0.025 & \cellcolor{cleanhitgreen!6.8}\textbf{0.068} \\
Control-flow selection & 17 & 17 & 16 & \cellcolor{cleanhitgreen!11.8}\textbf{0.118} & \cellcolor{cleanhitgreen!0.0}0.000 & \cellcolor{cleanhitgreen!0.0}0.000 & \cellcolor{cleanhitgreen!0.0}0.000 & \cellcolor{cleanhitgreen!0.0}0.000 & \cellcolor{cleanhitgreen!0.0}0.000 & \cellcolor{cleanhitgreen!5.9}0.059 & \cellcolor{cleanhitgreen!5.9}0.059 \\
\multicolumn{12}{@{}l}{\textbf{\emph{Applicability}}} \\
Precondition mismatch & 54 & 54 & 54 & \cellcolor{cleanhitgreen!11.1}0.111 & \cellcolor{cleanhitgreen!1.9}0.019 & \cellcolor{cleanhitgreen!5.6}0.056 & \cellcolor{cleanhitgreen!7.4}0.074 & \cellcolor{cleanhitgreen!0.0}0.000 & \cellcolor{cleanhitgreen!7.4}0.074 & \cellcolor{cleanhitgreen!13.0}\textbf{0.130} & \cellcolor{cleanhitgreen!11.1}0.111 \\
\multicolumn{12}{@{}l}{\textbf{\emph{Output behavior}}} \\
Representation schema & 43 & 41 & 24 & \cellcolor{cleanhitgreen!20.9}\textbf{0.209} & \cellcolor{cleanhitgreen!18.6}0.186 & \cellcolor{cleanhitgreen!9.3}0.093 & \cellcolor{cleanhitgreen!20.9}\textbf{0.209} & \cellcolor{cleanhitgreen!7.0}0.070 & \cellcolor{cleanhitgreen!7.0}0.070 & \cellcolor{cleanhitgreen!18.6}0.186 & \cellcolor{cleanhitgreen!20.9}\textbf{0.209} \\
Return / exception / side effect & 50 & 25 & 25 & \cellcolor{cleanhitgreen!0.0}0.000 & \cellcolor{cleanhitgreen!0.0}0.000 & \cellcolor{cleanhitgreen!2.0}0.020 & \cellcolor{cleanhitgreen!0.0}0.000 & \cellcolor{cleanhitgreen!0.0}0.000 & \cellcolor{cleanhitgreen!0.0}0.000 & \cellcolor{cleanhitgreen!4.0}0.040 & \cellcolor{cleanhitgreen!30.0}\textbf{0.300} \\
\multicolumn{12}{@{}l}{\textbf{\emph{Evidence role}}} \\
Aggregate vs. instance & 118 & 59 & 1 & \cellcolor{cleanhitgreen!44.9}0.449 & \cellcolor{cleanhitgreen!36.4}0.364 & \cellcolor{cleanhitgreen!72.0}0.720 & \cellcolor{cleanhitgreen!55.1}0.551 & \cellcolor{cleanhitgreen!73.7}\textbf{0.737} & \cellcolor{cleanhitgreen!67.8}0.678 & \cellcolor{cleanhitgreen!65.3}0.653 & \cellcolor{cleanhitgreen!58.5}0.585 \\
Intent vs. outcome & 196 & 98 & 3 & \cellcolor{cleanhitgreen!48.5}0.485 & \cellcolor{cleanhitgreen!87.2}0.872 & \cellcolor{cleanhitgreen!85.2}0.852 & \cellcolor{cleanhitgreen!84.2}0.842 & \cellcolor{cleanhitgreen!90.8}\textbf{0.908} & \cellcolor{cleanhitgreen!83.7}0.837 & \cellcolor{cleanhitgreen!90.3}0.903 & \cellcolor{cleanhitgreen!88.3}0.883 \\
Measurement vs. calibration & 74 & 37 & 1 & \cellcolor{cleanhitgreen!68.9}0.689 & \cellcolor{cleanhitgreen!91.9}0.919 & \cellcolor{cleanhitgreen!100.0}\textbf{1.000} & \cellcolor{cleanhitgreen!95.9}0.959 & \cellcolor{cleanhitgreen!98.6}0.986 & \cellcolor{cleanhitgreen!100.0}\textbf{1.000} & \cellcolor{cleanhitgreen!100.0}\textbf{1.000} & \cellcolor{cleanhitgreen!95.9}0.959 \\
Normative vs. empirical & 276 & 138 & 4 & \cellcolor{cleanhitgreen!72.5}0.725 & \cellcolor{cleanhitgreen!81.2}0.812 & \cellcolor{cleanhitgreen!88.8}0.888 & \cellcolor{cleanhitgreen!87.3}0.873 & \cellcolor{cleanhitgreen!87.7}0.877 & \cellcolor{cleanhitgreen!96.0}\textbf{0.960} & \cellcolor{cleanhitgreen!93.1}0.931 & \cellcolor{cleanhitgreen!83.7}0.837 \\
State change vs. rationale & 106 & 53 & 3 & \cellcolor{cleanhitgreen!100.0}\textbf{1.000} & \cellcolor{cleanhitgreen!79.2}0.792 & \cellcolor{cleanhitgreen!100.0}\textbf{1.000} & \cellcolor{cleanhitgreen!97.2}0.972 & \cellcolor{cleanhitgreen!99.1}0.991 & \cellcolor{cleanhitgreen!88.7}0.887 & \cellcolor{cleanhitgreen!100.0}\textbf{1.000} & \cellcolor{cleanhitgreen!79.2}0.792 \\

%% file: sections/5_analysis_limitations_working890.tex
\section{Discussion}
\label{sec:validity}

\paragraph{Why this benchmark matters.}
We built SameCapRisk-Bench to study a distinction that capability-level
relevance does not resolve: two skills may serve the same purpose, but only
one meets the current query's requirements. Each contrast makes that
distinction inspectable through a query requirement, two skill passages,
and supporting evidence. Paired queries further test whether the preferred
skill changes when the requirement changes. This design lets the community
evaluate query-specific compatibility alongside conventional helpful recall.

\paragraph{What the public retrievers reveal.}
The evaluated methods frequently retrieve a helpful skill together with
its conflicting sibling. On single-query contracts, the four specialized
retrievers expose the marked sibling in 0.737--0.881 of their top-3 lists;
ordering errors persist even when both skills are retrieved. Their much
lower exposure on source-role queries shows why a pooled score is not
enough to characterize this behavior. Reranking and longer inputs also
affect different parts of the problem: longer inputs improve the
first-place choice, but do little to remove the sibling from the list.
The benchmark therefore separates recovering the helpful skill, ordering
the two representatives, and returning a list without the marked conflict.

\paragraph{Resource and access.}
The public repository is
\url{https://github.com/jdding/SameCapRisk}.
The reviewer bundle contains versioned query and skill records, fixed
rankings, a CPU evaluator, and evidence indexes with selected checks.
Pinned reconstruction scripts restore omitted source texts and verify
their hashes. Appendix~\ref{app:review-files} describes what the bundle
covers and the separate code and data terms.

\paragraph{Using the resource.}
A new retriever can rank the label-free query and candidate texts in each
protocol's assigned pool, then submit a top-20 list per query to the CPU
evaluator. It checks the rankings and recomputes Recall, HSR, and CleanHit.
The protocol rows reveal whether a gain comes from source-task contracts,
evidence-role exchange, or output-interface exchange; the taxonomy rows
identify the particular requirement, such as a field binding or
precondition, behind a failure. Targeted evaluations should retain the
full query and unit counts, including failures, and state what information
the method received. A method using released family edges must be reported
separately from text-only retrievers.

\paragraph{Limitations.}
\label{sec:scope}
Our conclusions concern the retrieval stage: whether a fixed candidate
list contains a helpful skill and its marked, query-incompatible sibling.
They do not establish which skill an agent will select or what will happen
during tool or skill execution. This scope follows from the labels, which
are supported by local query--skill contract evidence rather than
whole-skill execution tests. HSR consequently measures exposure of the
annotated sibling, not the prevalence of all incompatible skills in a
registry. Within this scope, generalization is limited by how the contrasts
were constructed. Candidate-only predictability reveals residual wording
cues, and independent human inter-annotator agreement has not yet been
measured. The protocols also differ in source, pool, text length, and
construction; source-role cases cover only ten declared groups, while
output-interface cases form a small set of derived pairs. We therefore
interpret their scores as results for these task settings, rather than
isolated effects of conflict type.

%% file: sections/6_conclusion.tex
\section{Conclusion}

SameCapRisk-Bench tests whether retrieval distinguishes skills with the
same capability but different query contracts. Its contrasts link the
requirement, both relevant skill passages, and supporting evidence.
Public retrievers often find the helpful skill while retaining its
conflicting sibling. Single-query contracts remain difficult even when
both candidates are retrieved; reranking gains are concentrated in
source-role queries. These results separate candidate recovery, relative
ordering, and list composition instead of treating helpful recall as
the whole outcome.

Future retrieval methods should preserve the distinguishing contract in
their inputs and use it to select query-compatible representatives without
exposing conflicting siblings. We intend to extend the versioned benchmark
with more diverse, independently checked contrasts and to test such methods
under the same information and candidate-pool boundaries. We will also
investigate how retrieval exposure affects downstream agent selection
and task outcomes.

%% file: sections/A_appendix_working890.tex
\section{Benchmark Specification}
\label{app:benchmark-construction}

\subsection{Label Evidence and Data Splits}
\label{app:admission}

GPT-5.6 assisted with processing and organizing the benchmark data,
including query--skill record preparation and consistency checks.
The authors defined the admission rules and remain responsible for
the released data.

Label evidence depends on the query requirement: numerical parameters can
be checked by calculation, whereas source-role contrasts require comparing
the supplied and requested evidence. We distinguish these forms of support
so that static judgments are not mistaken for execution outcomes.
Counting each of the 890 units once by its
primary evidence gives 160 mathematical or algorithmic certificates,
37 interface-contract checks, 37 tool/backend or numerical outcome
contrasts, 387 static source-role comparisons, 241 source-contract or
local-operation reviews, and 28 code-change records with source references.

The certificates verify parameters, computations, or preconditions through
formula and set comparisons. The outcome contrasts compare observed
results: 15 attached-tool checks against
MedCalc answers, 10 ToolQA backend contrasts, and 12 TheoremQA numerical
contrasts. Interface checks assess the requested interface or permission
contract. Source-role comparisons assess whether the supplied evidence
has the role the query requires. The 241 source-contract or local-operation
reviews include 47 bounded operation checks and 51 source derivations,
alongside task/checker bindings and 2 helpful-side-only certificates.
Code-change records locate the modified region and its source. In each
case, the evidence supports the stated local contrast; historical execution
results apply to the implementation on which they were obtained.

Evidence must remain applicable after a skill is revised. Each judgment
therefore identifies the assessed query and skill passages and binds their
exact texts through hashes. Reused evidence links to its original record,
with check or proof-file hashes where available. A change to an assessed
passage requires repeating the affected check and updating that binding,
so evidence for an earlier skill cannot silently support a different one.

The split assignments preserve these relationships. The 55 single-query
and 24 output-interface development queries retain their construction-time
roles; the remaining queries form the main evaluation. Both queries of a
paired unit stay together, as do connected components formed by shared
source lineage or identical texts. Source-role pairs are evaluation-only.
These components also keep related cases together in the candidate-only
diagnostic. Construction diagnostics use the full benchmark, including
development cases, whereas retrieval scores use only the main split.
Taxonomy labels and split assignments were fixed before retriever evaluation.

\subsection{Distinguishing Conflict Types}
\label{app:taxonomy}

We classify a contrast by the smallest change that makes the conflicting
skill satisfy the query. This rule distinguishes the cause of a mismatch
from its later consequences. For example, passing a value into the wrong
input field is a field/parameter-binding conflict, even if it also produces
an incorrect output. Returning the wrong field set after an otherwise
appropriate computation is instead a representation/schema conflict.
Likewise, changing how a result is computed belongs to procedure; changing
only how that result is returned belongs to output behavior. We record
secondary consequences without assigning a second primary type. If two
independent corrections are necessary, the contrast does not meet the
single-conflict admission rule and is rejected or left unresolved.

\subsection{Inspecting Labels and Reproducing Scores}
\label{app:review-files}

The reviewer bundle supports label inspection and score reproduction.
An evidence index links every unit to its query and skill records and
supporting check excerpt, contract review, or code-change record.
The 9 support files cover all 890 units with selected excerpts rather
than complete original proof files.

For score reproduction, we provide fixed rankings and a CPU evaluator.
Some source texts are distributed through reconstruction scripts: these
restore 6,660 background skills and 59 CHAMP queries from pinned upstream
files and verify their hashes. Once the inputs are restored, the evaluator
checks query coverage, candidate-pool membership, rank uniqueness, and
input identity before computing the metrics. We also provide the rankings
and replay command for the longer-input reranker comparison. Our code and
the third-party query and skill data have separate licensing terms;
the data remain governed by their upstream licenses.

\section{Interpreting Retrieval Results}
\label{app:retrieval-analyses}

\subsection{How Task Composition Affects the Scores}
\label{app:weighting}
The pooled score reflects a particular task mixture. With equal weight
per query, the 411 single-query, 774 source-role, and 50 output-interface
queries give source-role performance the largest influence. Unit-macro
scores first average each unit's one or two queries, then weight the
823 units equally; group-macro scores weight declared groups equally.
All three summaries accompany the fixed rankings.

Reweighting tasks can change the leading method without changing any
prediction. When we vary the single-query and source-role weights, RRF
leads CleanHit at zero source-role weight; BGE reranking leads at weights
of 0.5 and above in the tested grid. Output-interface results are reported
separately. We retain the original mixture for the main result, while
this comparison shows why its leading method need not lead on another
task mix.

The source-role advantage also depends on which groups are included.
Removing each of its 10 groups in turn, with predictions held fixed,
changes the leading CleanHit@3 method in 3 source-role comparisons and
1 pooled comparison. These groups contain 2--122 queries. Removing the
entire source-role domain changes the pooled leader, whereas removing
any of the 8 smaller source domains leaves BGE reranking first. Thus,
the pooled ranking is most sensitive to the large source-role component.
For the reported intervals, we resample declared groups within each
protocol, sharing draws across methods and retaining the original query
weights. These intervals describe variation under the stated grouping
and task mixture.

\begin{table*}[t]
\caption{Helpful retrieval (Recall) and marked-sibling exposure (HSR) at
three list cutoffs on the 1,235 main queries. Higher Recall and lower HSR
are preferred. Rows follow the method order in Table~\ref{tab:protocol}.}
\label{tab:cutoff}
\centering
\small
\setlength{\tabcolsep}{4pt}
\begin{tabular*}{\textwidth}{@{}l@{\extracolsep{\fill}}rrrrrr@{}}
\toprule
& \multicolumn{2}{c}{$K=3$} & \multicolumn{2}{c}{$K=5$} & \multicolumn{2}{c}{$K=20$} \\
\cmidrule(lr){2-3}\cmidrule(lr){4-5}\cmidrule(l){6-7}
Method & Recall@3 $\uparrow$ & HSR@3 $\downarrow$ & Recall@5 $\uparrow$ & HSR@5 $\downarrow$ & Recall@20 $\uparrow$ & HSR@20 $\downarrow$ \\
\midrule
\input{generated/working890_cutoff_rows.tex} 
\bottomrule
\end{tabular*}
\end{table*}

\subsection{Separating Candidate Recovery from Ordering}

The same-candidate comparison measures what reranking changes after
retrieval has already determined which skills are available. BGE reranking
and Qwen listwise reranking both reorder the BGE top-20 candidates. Any
CleanHit difference from BGE dense ranking therefore comes from changes
within this fixed candidate set, rather than recovery of additional skills.
We calculate each query's change in binary CleanHit and resample these
paired differences by the declared groups. The protocol-specific gains
and intervals are shown in Figure~\ref{fig:matched-reranking}.

An additional comparison isolates the relative order of the two marked
skills. All 7 neural methods retrieve both on the same 543 main queries,
including 309 single-query cases. On those single-query cases, helpful-first
rates range from 0.521 to 0.696: even the best point estimate leaves about
30\% ordered the other way. BGE reranking has a lower helpful-first rate
than BGE dense ranking (0.521 versus 0.563), so reranking does not uniformly
improve this distinction. The subset covers 75.2\% of single-query cases
but only 24.3\% of source-role cases. It answers an ordering question when
both candidates are available, rather than replacing full-split retrieval
scores.

\subsection{Input Length and Output Handling}

Retrieving both skills does not guarantee that the ranker sees their
distinguishing text. The main BGE dense setting uses
512 document tokens; its 4,096-token run measures sensitivity to input
length. SkillRouter and SkillSight retain their released formatting and
budgets, while Qwen receives a 1,200-character skill excerpt. These are
the inputs under which the reported rankings were obtained.

To examine the reranker's input budget separately, we increase it from
512 to 4,096 tokens while fixing the BGE top-20 sets, model weights,
tokenizer, FP32 scoring, and tie rule. Only the batch size changes from
16 to 1 to fit memory; a 512-token check on 2 queries per pool reproduces
the earlier ordered lists. The longer input leaves all 774 source-role
lists unchanged. On single-query contracts, CleanHit@3 changes by
$+0.0073$ (15 gains, 12 losses; 95\% group-resample interval
[$-0.0164$, $0.0303$]); the output-interface change is $-0.02$
[$-0.08$, $0.04$]. Both intervals include zero. Together with the improved
first-place choice, this result shows that better top-ranked selection can
coexist with little change in whether the conflicting sibling remains in
the list. We retain the original 512-token setting in the main comparison.

Qwen also uses fixed output handling. Across all 1,314 cases, 27 JSON
lists require missing candidates to be appended, and 3 outputs use the
non-JSON numeric fallback. All cases remain in the denominator. Of the
incomplete lists, 26 occur in the main split; their appended candidates
are below rank 3. The scores thus include output-handling failures.

\subsection{Cutoff Sensitivity}

Table~\ref{tab:cutoff} shows how returning more candidates changes both
helpful retrieval and marked-sibling exposure. For BGE-M3, increasing $K$ from 3 to 20
raises Recall from 0.811 to 0.937, but HSR rises much more, from 0.329 to
0.833. More room in the list therefore recovers helpful skills without
necessarily making the list cleaner. The accompanying results provide all
three metrics at $K=1,3,5,20$. At $K=1$, CleanHit equals Recall, and HSR
counts queries where the sibling ranks first. Comparing this outcome with
longer lists separates an incorrect first choice from a conflicting skill
appearing farther down the list.

\section{Construction Diagnostics}
\label{app:construct-validity}

\subsection{Candidate-Only Predictability}

The candidate-only diagnostic asks whether skill wording reveals the
label without the query. We use
five group-disjoint folds and character-within-word
TF--IDF (3--5 grams, sublinear term frequency, at most 30,000 features)
with logistic regression ($C=1$). Both are fitted only on training
candidates in each fold. We select the higher-scored candidate within each
held-out pair and give ties half credit. The 59.66\% accuracy on 466
single-query units, with higher scores on control-flow and schema contrasts, shows that
candidate wording still provides uneven label cues across conflict types.

\subsection{Query Dependence and Candidate Order}

The query-conditioned probe uses \texttt{Qwen3-4B-Instruct-2507} at the
fixed revision recorded in its run metadata, with a 32,768-token context
limit and at most 192 new tokens. It receives the complete query and both
skill texts under neutral A/B identifiers and uses greedy decoding.
The response schema permits A, B, or \texttt{UNDECIDABLE}. We run
1,314 original-query jobs, 1,115 source/type-mismatched-query jobs, and
64 mirrored-order jobs, giving 2,493 outputs. We compute original-query
accuracy as the number of valid correct selections divided by all
original-query jobs, including abstentions and invalid responses.
All 148 invalid outputs remain in
their respective denominators.

The mirrored-order test keeps the query fixed and exchanges the candidate
positions. The probe chooses the compatible skill in both presentations
on 43/64 selected cases. The query-mismatch test instead keeps both skills
fixed and replaces the query with another from the same source and conflict
type. On the 1,115 cases with an eligible replacement, the rate of selecting
the original query's helpful skill falls from 0.836 to 0.140. Both rates
refer to the original label: we measure whether changing the query changes
the choice, rather than judging which skill fits the replacement query.
Together, the tests distinguish sensitivity to the requested task from
sensitivity to candidate position.

\subsection{Probe Response Validity}
\label{app:probe-outcomes}

We count a probe response as correct only when it is valid and selects the
labelled helpful skill. Selecting the alternative, abstaining, or returning
malformed output all count as unsuccessful, but we report these outcomes
separately to distinguish selection errors from response-format failures.
The 466 single-query jobs yield 298
target-consistent choices, 77 alternative choices, 28 abstentions, and
63 invalid outputs. The 774 source-role jobs yield 765, 8, 1, and 0,
respectively; the 74 output-interface jobs yield 65, 7, 0, and 2.
Each set of counts covers all original-query jobs in that protocol.

The 63 invalid single-query outputs all reach the 192-token generation
limit with an unterminated JSON string. Their inputs contain 329--18,706
tokens and fit within the 32,768-token context budget: in these cases,
incomplete output, rather than an overlength input, prevents a valid
selection. The fixed parser keeps these outputs invalid; we do not infer
intended answers.

\subsection{Cross-Family Model-Based Label Audit}
\label{app:cross-family-audit}

We used the DeepSeek-V4-Pro
and GLM-5.3 APIs on 64 single-query units, 16 source-role pairs, and
16 output-interface pairs. These 96 held-out units contain 128 queries
from 90 lineage groups, selected to cover different conflict types and
source groups.

Each API request contains the complete query and both skill texts under
neutral A/B identifiers. We withhold the recorded labels, construction
evidence, and the other model's answer. For each skill, the model judges
whether it meets the query's requirement and quotes the relevant passage.
It then selects one skill, accepts both, rejects both, or reports that the
supplied text is insufficient to decide. We check the quotations against
the input text. Repeating each query with A and B exchanged gives
256 judgments per model.

The models agree on 124/128 original-order queries ($\kappa=0.94$).
Exchanging candidate positions preserves DeepSeek's judgment on 123/128
queries and GLM's on 126/128. On 88/96 units, both models select the
recorded helpful skill for every query in both orders.

We reviewed the remaining 8 units against their query and skill texts.
We found construction defects in 2 units: one query asked for a weight count,
but the helpful skill returned an activation count; another required 3 answer
fields, but the helpful skill specified 4. Review supported the recorded
labels in the other 6 units. We then checked the same
source and template families across the 890-unit roster. This found 2
more field-count mismatches and 1 skill that listed possible positive
root counts when the query required the actual count. We corrected these
5 units in a separate data revision. The agreement figures above describe
the audit before these repairs; the retrieval tables use the corrected texts.

%% file: generated/working890_cutoff_rows.tex
RRF & 0.781 & 0.358 & 0.859 & 0.488 & 0.951 & 0.857 \\
BGE-M3 & 0.811 & 0.329 & 0.867 & 0.442 & 0.937 & 0.833 \\
\midrule
SkillRet & 0.903 & 0.344 & 0.921 & 0.445 & 0.964 & 0.749 \\
SkillSight & 0.926 & 0.388 & 0.964 & 0.495 & 0.989 & 0.802 \\
SkillRouter & 0.939 & 0.386 & 0.955 & 0.446 & 0.968 & 0.590 \\
R3-Skill & 0.944 & 0.393 & 0.967 & 0.465 & 0.987 & 0.577 \\
\midrule
BGE reranker & 0.863 & 0.297 & 0.891 & 0.388 & 0.937 & 0.833 \\
Qwen2.5-7B & 0.849 & 0.324 & 0.876 & 0.415 & 0.937 & 0.833 \\